\documentclass[conference]{IEEEtran}
\IEEEoverridecommandlockouts
\usepackage[dvips]{graphicx}
\usepackage[english]{babel}
\usepackage{algorithm}
\usepackage{algorithmic}
\usepackage{amsfonts}
\usepackage{amsmath,amssymb}
\usepackage{booktabs}
\usepackage[numbers,sort&compress]{natbib}
\usepackage{bm}
\usepackage{enumerate}
\usepackage{amsfonts,amssymb,amsmath}
\usepackage{multirow}
\usepackage{epstopdf}
\usepackage{stfloats}
\usepackage{caption}
\usepackage{url}
\usepackage{psfrag}
\newtheorem{definition}{\bf{Definition}}

\begin{document}

\title{Deployment of 5G Networking Infrastructure with Machine Type Communication Considerations \vspace{-0.5cm}}
\author{\IEEEauthorblockN{Xiangxiang Xu$^\dagger$, Walid Saad$^*$, Xiujun Zhang$^\dagger$, Limin Xiao$^\dagger$ and Shidong Zhou$^\dagger$} \thanks{This work was supported by National Basic Research Program of China (Grant No. 2012CB316002), National Natural Science Foundation of China (Grant No. 61201192), Science Fund for Creative Research Groups of NSFC (61321061), National S\&T Major Project (Grant No. 2015ZX03001016-002), National High Technology Research and Development Program of China (863 Program) (Grant No. 2014AA01A704), Tsinghua University Initiative Scientific Research (2015Z02-3), International Science and Technology Cooperation Program (Grant No. 2012DFG12010), Open Research Fund of National Mobile Communications Research Laboratory, Southeast University (Grant No. 2012D02), Tsinghua-Qualcomm Joint Research Program, the Tsinghua- Intel International S\&T Cooperation Program (ICRI-MNC), and the U.S. Office of Naval Research (ONR) under Grant N00014-15-1-2709. }
\IEEEauthorblockA{\small $^\dagger$ State Key Laboratory on Microwave and Digital Communications,\\
Tsinghua National Laboratory for Information Science and Technology,\\
Department of Electronic Engineering, Tsinghua University, Beijing 100084, China\\
$^*$ Wireless@VT, Bradley Department of Electrical and Computer Engineering, Blacksburg, VA, USA.\\
Emails: xuxxmail@163.com, walids@vt.edu, zhangxiujun@tsinghua.edu.cn, xiaolm@mail.tsinghua.edu.cn, zhousd@mail.tsinghua.edu.cn}\vspace{-1cm}}

\maketitle

\begin{abstract}
Designing optimal strategies to deploy small cell stations is crucial to meet the quality-of-service requirements in next-generation cellular networks with constrained deployment costs. In this paper, a general deployment framework is proposed to jointly optimize the locations of backhaul aggregate nodes, small base stations, machine aggregators, and multi-hop wireless backhaul links to accommodate both human-type and machine-type communications. The goal is to provide deployment solutions with best coverage performance under cost constraints. The formulated problem is shown to be a multi-objective integer programming for which it is challenging to obtain the optimal solutions. To solve the problem, a heuristic algorithm is proposed by combining Lagrangian relaxation, the weighted sum method, the $\epsilon$-constraint method and tabu search to obtain both the solutions and bounds, for the objective function. Simulation results show that the proposed framework can provide solutions with better performance compared with conventional deployment models in scenarios where available fiber connections are scarce. Furthermore, the gap between obtained solutions and the lower bounds is quite tight.
\end{abstract}

\IEEEpeerreviewmaketitle
\section{Introduction}

In next-generation cellular networks, low power small base stations (SBSs) will be densely deployed to boost the coverage and capacity \cite{Andrews14,Senza13}. Furthermore, billions of machines will coexist with conventional human-type communication (HTC) \cite{Zhou13,Yaacoub14}. Thus, next-generation cellular systems must integrate heterogeneous base stations and heterogeneous types of communications. All these components should be considered in the network planning period to obtain a network deployment strategy with optimal operating performance. Indeed, one major challenge for next-generation cellular networks is to design flexible, scalable and low cost deployment strategies \cite{Senza13} that capture the network's heterogeneity.

For SBSs, backhaul links are required to connect them to the core network. The performance of an SBS also depends on the capacity of the backhaul link. Thus, backhaul links and SBSs should be considered jointly during network planning \cite{Senza13} to guarantee the system performance. Conventional works on base station (BS) deployment have focused mainly on scenarios in which fiber connection backhaul links are available for all candidate locations \cite{Richter12,Brevis11,Ghazzai15,Hsieh14,Zhao14}.
%
In other words, only sites with existing fiber connections are regarded as potential sites. However, as discussed in \cite{Senza13}, the availability of existing fiber connections might be scarce which would hamper the flexibility and  performance of the network. Also, deploying new fiber connections for dense SBSs is unacceptable from the perspective of cost. To overcome this problem, the works in \cite{Senza13,Singh14} have shown that heterogeneous backhaul, wired and wireless, would be a viable solution for ultra dense small cell network. The authors in \cite{Islam14} studied the deployment of wireless backhaul links for SBSs. However, this work only studied the placement of wireless backhaul nodes which provide backhaul links to SBSs directly, through one hop, while SBSs deployment has been neglected.

Although interesting deployment problems have been studied, existing works \cite{Richter12,Brevis11,Ghazzai15,Hsieh14,Zhao14,Islam14} have only focused on fibre backhaul links and HTC. Such works can not be extended to solve the deployment problem of the next-generation networks since the impact of MTC and the designing of backhaul network have been neglected. In particular, heterogeneous communications, HTC and machine-type communication (MTC), and heterogeneous backhaul links, wired and wireless,  should be considered jointly in the planning period. In addition, multi-hop wireless backhaul links should be allowed to improve the network performance when the available fiber links are scarce. To our best knowledge, the general framework of the deployment of next-generation network while taking into account jointly HTC, MTC, wired backhaul and wireless backhaul, has never been addressed yet.

\begin{figure}[t]
\centering
\scalebox{0.40}{\includegraphics*[0,0][368,275]{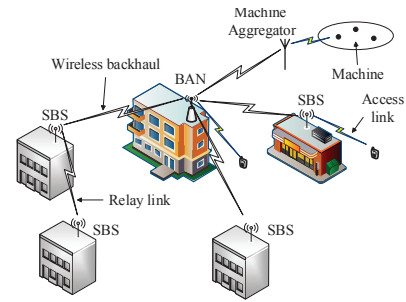}}
\caption{System model.}
\label{Topology}
\vspace{-20pt}
\end{figure}
The main contribution of this paper is to propose a practical general deployment framework which incorporates HTC, MTC, available wired backhaul links and multi-hop wireless backhaul links for next-generation cellular networks. Three types of stations, backhaul aggregate nodes (BANs), small base stations (SBSs) and machine aggregators (MAs), are deployed to get the network with optimal performance. In particular, MAs are responsible for covering MTC and backhaul links of an SBS can be provided either by some BAN or by other SBSs. We formulate the deployment problem as a multi-objective integer programming which is NP-hard. To obtain the entire Pareto optimal solution set, we propose an tabu search based adaptive algorithm which can also provide lower bounds for the problem and the bounds can be used to check the quality of the obtained solutions. Simulation results show that our framework can get good solutions in scenarios where available fiber connections are scarce. Furthermore, the ratio of the objective value of the obtained solutions to the lower bounds can be as tight as 1.99.

The rest of the paper is organized as follows. Section II presents the system and problem formulation. The proposed deployment algorithm is introduced and discussed in Section III. Section IV analyzes the simulation results and conclusions are drawn in Section V.

\section{System model and problem formulation}
Our model is shown in Fig. \ref{Topology}. The backhaul aggregate nodes (BANs) are placed at sites with available fiber connections to connect SBSs to the core network. Meanwhile, the deployed SBSs with backhaul can act as relays and connect other SBSs to BANs. Thus, for an SBS, the backhaul link can be provided either by a BAN or by other SBSs.
In addition, we assume that BANs have access ability and can cover a certain area.

For MTC, a clustered infrastructure \cite{Park15,Yaacoub14} is adopted in which the machines first transmit data to an machine aggregator (MA)  which will then send the data to a BAN as shown in Fig. \ref{Topology}. In the system, we mainly consider static machines whose locations and data rates are assumed to be  known.
 Unlike HTC, for MTC, uplink transmission is our focus in the deployment period.

For HTC, both access links and backhaul links are  assumed to work in the orthogonal millimeter-wave (mmW) bands.
 For MTC, a conventional, sub-6 GHz band is used for access since MTCs require long range, low rate coverage, which is in contrast to mmW which is a high bandwidth, short-range solution.
For mmW propagation, the pathloss model in \cite{Ghosh14} is used:
\begin{align}
Loss(d)=20{\log _{10}}\left(\frac{{4\pi {d_0}}}{\lambda }\right){\rm{ + }}10\bar n{\log _{10}}\left(\frac{d}{{{d_0}}}\right) + \chi, \label{PropagationModel}
\end{align}
where $d_0$ is a reference distance and $\lambda$ is the wavelength of the carrier. $\bar n$ denotes the path loss exponent and $d$ is transmission distance. $\chi$ is the shadowing component which is a zero mean Gaussian variable with a standard deviation ${\sigma}_s$ in dB. The value $\bar n$ and ${\sigma}_s$ depend on the transmission conditions, i.e., line of sight (LOS) or non line of sight (NLOS), access link or backhaul link. The LOS probability derived in \cite{Kulkarni14} is utilized in our work, $p_{LOS}\left(d\right)=e^{-{\beta}d}$, where $d$ denotes transmission distance and $\beta$ is related with transmission environment.
%

\subsection{Problem formulation}
Consider an area $\cal A$ in which a number of  BANs, SBSs and MAs must be deployed to serve both HTC and MTC. In this area, the potential sites set is given by ${\cal F}={{\cal F}_1} \cup {{\cal F}_2} \cup {{\cal F}_3}$. Set ${{\cal F}_1}=\{1,\ldots,F_1\}$ indicates sites with available fiber connection and denotes the candidate locations for BANs. Set ${{\cal F}_2}=\{1,\ldots,F_2\}$ and ${{\cal F}_3}=\{1,\ldots,F_3\}$ denote the candidate locations for SBSs and MAs, respectively.

We define the deployment variables ${z_k} \in \{0,1\}$, ${y_i} \in \{0,1\}$ and ${w_j} \in \{0,1\}$ for BANs, SBSs and MAs, respectively, to indicate the station is deployed ($=1$) or not ($=0$).
For HTC, in order to indicate the area coverage, we partition the area $\cal A$ equally into a set of subareas which are elements of the set ${\cal S}=\{1,\ldots,S\}$. For each subarea $s \in {\cal S}$, the center point is chosen to indicate its coverage.
 Variables $x_{ks}$ and $x_{is}$ are defined as follows to indicate the coverage of subarea $s$ by BAN $k$ and SBS $i$, respectively.
\begin{align*}
 x_{ks}\textrm{ }(x_{is}) = &\left\{ \begin{array}{l}
 1 \quad \textrm{if subarea $s$ is covered by BAN $k$ (SBS $i$).}\\
 0 \quad \textrm{otherwise.}\\
 \end{array} \right.
\end{align*}
As previously mentioned, we assume that SBSs can act as relays.
 To bound the transmission latency, it is reasonable to restrict the number of hops from an SBS to a BAN. Here, $N$ is used to denote the maximum number of relays for a connection from an SBS to a BAN. Then, the maximum number of hops from an SBS to a BAN is $N+1$. For the multi-level network, we define the following variables to indicate the connections among BSs.
\begin{align*}
 x_{ki}\textrm{ }(x_{pi}) = &\left\{ \begin{array}{l}
 1 \quad \textrm{if BAN $k$ (SBS $p$) provides backhaul to SC $i$.}\\
 0 \quad \textrm{otherwise.}\\
 \end{array} \right.
\end{align*}
Here, we let $x_{ii}=0$  for all $i \in {\cal F}_2$. It needs to be clarified that
the level of our network infrastructure is not fixed. The maximum level is $N+1$ and some SBSs can connect to BANs through just one or two hops. For subarea $s$, we define the \emph{downlink} routing variables ${x_{ki}^s} \in \{0,1\}$ (${x_{pi}^s} \in \{0,1\}$) to indicate whether the data for $s$ is routed from BAN $k$ (SBS $p$) to SBS $i$ or not.

For MTC, we let $\mathcal{M} =\{1,\ldots,M\}$ be the set of the machines that need to be covered by MAs. Each machine $m \in {\cal M}$ has a data rate $d_m$. We define variables $x_{jm} \in \{0,1\}$ and $x_{kj} \in \{0,1\}$ to indicate the coverage of machine $m \in {\cal M}$ by MA $j \in {{\cal F}_3}$ and the availability of a link between BAN $k$ and MA $j$, respectively.

Given the area $\cal A$, our purpose is to find an optimal deployment solution which indicates the locations of BANs, SBSs, MAs and their connections. From an operator's perspective, we want to maximize the coverage with a minimum deployment cost. In order to examine the relationship between deployment cost and performance deeply, unlike the previous studies \cite{Richter12,Brevis11,Ghazzai15,Hsieh14,Zhao14}, we consider multiple objectives simultaneously and give an optimal deployment solution set from which the proper strategy can be chosen. The three objective functions are defined as follows.
\begin{align}
    &{f_1}\left(\bm{z},\bm{y},\bm{w}\right) =\sum\limits_{{k} \in {{\cal F}_1}} {{z_{{k}}}{c_{{k}}}}+ \sum\limits_{{i} \in {{\cal F}_2}} {{y_{{i}}}{c_{{i}}}}+\sum\limits_{{j} \in {{\cal F}_3}} {{w_{{j}}}{c_{{j}}}} ,\label{F1}\\
    &{f_2}\left({\bm{x}}\right) =S- \sum\limits_{s \in {\cal S}} {\sum\limits_{{k} \in {\cal F}_1} {{x_{ks}}} }-\sum\limits_{s \in {\cal S}} {\sum\limits_{{i} \in {\cal F}_2} {{x_{is}}} } , \\
    &{f_3}\left({\bm{x}}\right) = M-\sum\limits_{m \in {\cal M}} {\sum\limits_{{j} \in {\cal F}_3} {{x_{jm}}} },
\end{align}
where $f_1\left(\bm{z},\bm{y},\bm{w}\right)$ represents deployment cost and $f_2\left({\bm{x}}\right)$, $f_3\left({\bm{x}}\right)$ indicate the number of uncovered subareas and machines. $c_k$, $c_i$ and $c_j$ denote deployment cost for BAN $k$, SC $i$ and MA $j$, respectively. Vector 
$\bm{x}=\left[x_{ki},x_{ks},x_{is},x_{ki}^s,x_{ip}^s,x_{jm}\right]$, $\bm{y}=\left[y_i\right]$, $\bm{z}=\left[z_k\right]$ and $\bm{w}=\left[w_j\right]$ are the binary variables.

In the considered area $\cal A$, mobile users are assumed to be distributed according to a homogeneous Poisson Point Process (PPP) of density ${\lambda}_u$. A subarea is said to be covered if the outage probability is less than a predefined threshold, $p_{oa}$. Then, the deployment problem can be formulated as a multi-level multi-objective integer programming problem:
\begin{equation}
\textrm{(P)}\qquad \mathop {\min }\limits_{{\bm{x}},{\bm{y}},{\bm{z}},{\bm w}} \quad \left[{f_1}\left(\bm{y},\bm{z},\bm{w}\right),{f_2}\left({\bm{x}}\right),{f_3}\left({\bm{x}}\right)\right],  \label{Problem}
\nonumber
\end{equation}
\vspace{-15pt}
\begin{align}
\textrm{s.t.}\quad &{x_{ks}} \le {z_k} \qquad s \in {\cal S},k \in {\cal F}_1, \label{OpenToConnectAccess1} \\
&{x_{is}} \le {y_i} \qquad s \in {\cal S},i \in {\cal F}_2, \label{OpenToConnectAccess2}  \\
&x_{ki} \le y_i \qquad k \in {\cal F}_1, i \in {\cal{F}}_2, \label{OpenToConnectBackhaul1} \\
&x_{ki} \le z_k \qquad k \in {\cal F}_1, i \in {\cal{F}}_2, \label{OpenToConnectBackhaul2} \\
&x_{pi} \le y_{p} \qquad i,p \in {{\cal F}_2}, \label{OpenToServeSC} \\
&x_{pi} \le y_i \qquad i,p \in {{\cal F}_2}, \label{OpenToBeServedSC} \\
&x_{ki}^s \le x_{ki} \qquad s \in {\cal S}, k \in {{\cal F}_1}, i \in {{\cal F}_2}, \label{ConnectToRelayBAN} \\
&x_{ip}^s \le x_{ip} \qquad  s \in {\cal S}, i,p \in {{\cal F}_2}, \label{ConnectToRelaySC} \\
&\sum\limits_{k \in {\cal F}_1} {x_{ks}} + \sum\limits_{i \in {\cal F}_2} {x_{is}}\leq 1 \qquad s \in {\cal S}, \label{AtMostOneAccess}\\
&\sum\limits_{k \in {\cal F}_1} {{x_{ks}}p\left({\gamma}_{ks} \le {\gamma}_t\right)}+\sum\limits_{i \in {\cal F}_2} {{x_{is}}p\left({\gamma}_{is} \le {\gamma}_t\right)} \le p_{oa}, \label{SNRAccess} \\
&\sum\limits_{i \in {{\cal F}_2}} {{x_{ki}}} + \sum\limits_{j \in {{\cal F}_3}}{x_{kj}}  \le {N_b} \qquad {k} \in {{\cal F}_1}, \label{BackCapability} \\
&\sum\limits_{k \in {\cal F}_1}{x_{ki}}+{\sum\limits_{p \in {\cal F}_2}{x_{pi}}}=y_i   \qquad i \in {\cal F}_2, \label{OpenMustServedBackhaul} \\
&\sum\limits_{k \in {{\cal F}_1}}{\sum\limits_{i \in {{\cal F}_2}}{x_{ki}^s}}=\sum\limits_{i \in {{\cal F}_2}}{x_{is}} \qquad s\in {\cal S}, \label{SubareaMustToBAN} \\
&{x_{is}}+{\sum\limits_{p \in {{\cal F}_2}}{x_{ip}^s}}={\sum\limits_{k \in {{\cal F}_1}}{x_{ki}^s}}+{\sum\limits_{p\in {{\cal F}_2}}{x_{pi}^s}},  \label{RoutingEquality} \\
&\sum\limits_{p\in {{\cal F}_2}}{\sum\limits_{i \in {{\cal F}_2}}{x_{pi}^s}}+\sum\limits_{k \in {{\cal F}_1}}{\sum\limits_{i\in {{\cal F}_2}}{x_{ki}^s}} \le N+1 \qquad s\in {\cal S}, \label{HopConstraint} \\
&\sum\limits_{p \in {{\cal F}_2}}{x_{pi}p(r_i > C_{pi})}+\sum\limits_{k \in {{\cal F}_1}}{x_{ki}p(r_i>C_{ki})} \le p_{ob}, \label{BackhaulCapacityConstraint} \\
&x_{kj} \le z_k  \qquad k \in {{\cal F}_1}, j \in {{\cal F}_3}, \label{MachineBAN} \\
&\sum\limits_{j \in {{\cal F}_3}}{x_{jm}} \le 1 \qquad m \in {\cal M}, \label{MachineAtMostOneAccess}\\
&x_{jm}d_{jm} \le D_{jt} \qquad j \in {{\cal F}_3}, m \in {\cal M}, \label{MachineDistanceConstraint} \\
&{\tau}_j \le {\tau}_{th} \qquad j \in {{\cal F}_3}, \label{MachineLatencyConstraints} \\
&\sum\limits_{k \in {{\cal F}_1}}{x_{kj}}=w_j \qquad j \in {{\cal F}_3}, \label{MachineOpenMustServed} \\
&\sum\limits_{m \in {\cal M}}{x_{jm}d_m}\delta \le C_{kj} \qquad k \in {{\cal F}_1}, j \in {{\cal F}_3}. \label{MachineBackhaulCapacity}
\end{align}

Here, ${\gamma}_{ks}$ (${\gamma}_{is}$) denotes the received signal to noise ratio (SNR) at the center point of subarea $s$ from BAN $k$ (SBS $i$). ${\gamma}_t$ is an outage threshold for user access. $N_b$ is the maximum number of nodes that a BAN can serve. The random variable $r_i$ indicates the total data request from the covered area of SBS $i$. $C_{pi}$ ($C_{ki}$) is the capacity of the link from SBS $p$ (BAN $k$) to BS $i$. $p_{ob}$ is the maximum allowable probability with which the limited backhaul capacity induces large transmission latency. $d_{jm}$ is the distance from MA $j$ to machine $m$ and $D_{jt}$ indicates the coverage range of MA $j$. ${\tau}_{th}$ denotes the transmission latency threshold of MTC. $0< \delta \le 1$ represents the data compression ratio.

In problem (P), constraints (\ref{OpenToConnectAccess1}), (\ref{OpenToConnectAccess2}), (\ref{OpenToConnectBackhaul1}), (\ref{OpenToConnectBackhaul2}), (\ref{OpenToServeSC}), (\ref{OpenToBeServedSC}) and (\ref{MachineBAN}) mean that only deployed BSs can provide connections to users, other BSs and machines. (\ref{ConnectToRelayBAN}) and (\ref{ConnectToRelaySC}) guarantee that a connection is required to route data among SBSs. (\ref{AtMostOneAccess}) and (\ref{SNRAccess}) ensure that each subarea can be covered at most by one BS with a lower outage probability than the predefined threshold. Since millimeter wave band networks are noise limited \cite{Ghosh14,Singh14}, SNR is used to indicate the quality of service. Constraint (\ref{BackCapability}) indicates the backhaul capability of a BAN.
 Constraint (\ref{OpenMustServedBackhaul}) ensures that a \emph{deployed} SBS ($y_i=1$) must have backhaul links either through connecting to BANs or other SBSs. Also, we do not consider the case in which one SBS has access to multiple BANs or SBSs simultaneously. For subarea $s$, constraint (\ref{SubareaMustToBAN}) implies that the data for $s$ must be transmitted from a BAN to some SBS once it is covered by an SBS ($\sum\limits_{i \in {{\cal F}_2}}{x_{is}}=1$). (\ref{RoutingEquality})
means that the data for subarea $s$ can be transmitted to SBS $i$ when subarea $s$ is covered by SBS $i$ ($x_{is}=1$) or SBS $i$ relays data to some other SBS ($\sum\limits_{p \in {{\cal F}_2}}{x_{ip}^s}=1$).
 Constraint (\ref{HopConstraint}) is the constraint on the number of hops. (\ref{BackhaulCapacityConstraint}) characterizes the wireless backhaul capacity limits. Here, we mainly focus on the limited wireless backhaul capacity and assume that access capacity is sufficient.

 Constraint (\ref{MachineAtMostOneAccess}) means that the machine can be covered by at most one MA and (\ref{MachineDistanceConstraint}) is the transmission distance constraint for machines. (\ref{MachineLatencyConstraints}) ensures the mean uplink transmission delay of machines covered by MA $j$ is less than the predefined threshold. Constraint (\ref{MachineOpenMustServed}) requires that, for deployed MAs, some deployed BANs should provide wireless backhaul links. (\ref{MachineBackhaulCapacity}) implies that the backhaul capacity of an MA is sufficient to satisfy the aggregate machine demand. Here, we assume that the data rate of an machine is quite low and it can be satisfied by the connected MA. In problem (P), the hop limits and constraints (\ref{BackhaulCapacityConstraint}) on the large latency probability together guarantee that a long path does not exist in the deployed networks.

Based on channel model (\ref{PropagationModel}), constraint (\ref{SNRAccess}) can be transformed into the following constraint on the transmission distance.
\begin{align}
\sum\limits_{i \in {{\cal F}_2}}{x_{is}\left(d_{is}-D_{it}\right)}+\sum\limits_{k \in {{\cal F}_1}}{x_{ks}\left(d_{ks}-D_{kt}\right)} \le 0. \label{DistanceConstraint}
\end{align}
$d_{is} \left(d_{ks}\right)$ is defined as the distance from SBS $i$ (BAN $k$) to the center point of subarea $s$. $D_{it}$ and $D_{kt}$ denote the coverage range of SBS $i$ and BAN $k$, respectively.

Since users are distributed according to a PPP, (\ref{BackhaulCapacityConstraint}) can be transformed further to the following constraints on the number of covered subareas given the users' rate demand distribution.
\begin{align}
\sum\limits_{s \in {\cal S}}{x_{is}}+\sum\limits_{s \in {\cal S}}{\sum\limits_{p \in {{\cal F}_2}}x_{ip}^s} \le N_i\left(\bm{x}\right) \qquad i \in {{\cal F}_2}. \label{NewBackhaulCapacity}
\end{align}

Here,$N_i\left(\bm{x}\right)=\sum\limits_{k \in {{\cal F}_1}}{x_{ki}N_{ki}}+\sum\limits_{p \in {\cal F}_2}{{x_{pi}}N_{pi}}.$
 $N_{ki}$ ($N_{pi}$) denotes the number of subareas that SBS $i$ can cover when connected to BAN $k$ (SBS $p$). In \cite{Zhou13}, the author proposed a framework to compute the mean transmission latency ${\tau}_j$ for MA $j$ in MTC which is a function of the number of machines and resources.
Then, given available resources and latency requirement ${\tau}_{th}$, the method proposed in \cite{Zhou13} can be modified to obtain the following constraint on the total number of machines that an MA can cover.
\begin{align}
\sum\limits_{m \in {\cal M}}{x_{jm}} \le N_j w_j\qquad j \in {{\cal F}_3}. \label{NewLatencyConstraint}
\end{align}

Problem (P) is a \emph{multi-objective multi-level capacitated facility location problem (FLP)}. This problem is an extension of multi-level FLP which is known to be NP-hard \cite{Gabor10}.
Here, Pareto optimality is considered for the multi-objective programming.
\begin{definition}
  For a minimization programming with objective function ${\bm{f}}(\bm{x})=\left[f_1(\bm{x}),\dots,f_N(\bm{x})\right]$. A solution ${\bm{x}}^*$ is \emph{efficient} if there does not exist other solution $\bm{x}$ such that $f_i(\bm{x})\le f_i({\bm{x}^*}),\forall i$ and $f_i(\bm{x})< f_i({\bm{x}^*}),\exists i$. The point $\bm{f}({\bm{x}}^*)$ in the objective space is called \emph{Pareto optimal} or \emph{nondominated} if ${\bm{x}}^*$ is efficient.
\end{definition}

Hereinafter, we use efficient and Pareto optimal interchangeably.
 Our purpose is to find the optimal solution set and guide the practical deployment of BANs, SBSs and MAs.

\section{Proposed algorithm}
To solve problem (P), we first combine the weighted sum method with the $\epsilon$-constraint method \cite{Kirlik14}
 together to transform it to a single objective optimization problem. In the weighted sum method, we assume that the network planner has a parameter $\theta$ that captures the tradeoff between the HTC coverage and MTC coverage. Then, the coverage objective function can be expressed as
$f_c\left({\bm{x}}\right)=f_2\left({\bm{x}}\right)+\theta f_3\left({\bm{x}}\right)$.
Then, through $\epsilon$-constraint method, the deployment cost ${f_1}\left(\bm{y},\bm{z},\bm{w}\right)$ is taken as a constraint. 
Given $\theta$ and $\epsilon$, the problem is transformed to the following problem.
\begin{equation}
\left(\textrm{P}_{\epsilon}\right)\qquad \mathop {\min } f_c\left({\bm{x}}\right), \nonumber
\end{equation}
\vspace{-20pt}
\begin{align}
\textrm{s.t.} \quad &{f_1}\left(\bm{z},\bm{y},\bm{w}\right) \le {\epsilon}, \label{CostConstraints} \\
&(\ref{OpenToConnectAccess1}-\ref{AtMostOneAccess}),(\ref{BackCapability}-\ref{HopConstraint}),(\ref{MachineBAN}-\ref{MachineDistanceConstraint}),
(\ref{MachineOpenMustServed}-\ref{NewLatencyConstraint}). \nonumber
\end{align}

The number of binary variables in problem ($\textrm{P}_{\epsilon}$) is considerably large. Conventional algorithm like branch-and-bound algorithm and linear relaxation can not address the problem efficiently. Here, we adopt a scheme that integrates Lagrangian relaxation and tabu search together to solve $\left(\textrm{P}_{\epsilon}\right)$.
\subsection{Lagrangian relaxation}
We note that the backhaul capacity constraint (\ref{NewBackhaulCapacity}) makes the problem challenging to handle. So we introduce Lagrangian multipliers $\bm{\lambda}=\{{\lambda}_i\ge0,i \in {\cal F}_2\}$ and relax constraint (\ref{NewBackhaulCapacity}). Then, the relaxed problem ($\textrm{P}_{\bm{\lambda}}^{\epsilon}$) is obtained.
\begin{equation}
\left(\textrm{P}_{\bm{\lambda}}^{\epsilon}\right)\quad \mathop {\min } S{\rm{ + }}\theta M - \sum\limits_{k \in {{\cal F}_1}} {{m_k}}  + \sum\limits_{i \in {{\cal F}_2}} {{n_i}}  - \theta \sum\limits_{j \in {{\cal F}_3}} {\sum\limits_{m \in {\cal M}} {{x_{jm}}} }, \nonumber
\end{equation}
\vspace{-10pt}
\begin{align}
\textrm{s.t. }
&(\ref{OpenToConnectAccess1}-\ref{AtMostOneAccess}),(\ref{BackCapability}-\ref{HopConstraint}),(\ref{MachineBAN}-\ref{MachineDistanceConstraint}),
(\ref{MachineOpenMustServed}-\ref{DistanceConstraint}), (\ref{NewLatencyConstraint}),(\ref{CostConstraints}), \nonumber
\end{align}
where
\begin{align*}
{m_k} &= \sum\limits_{s \in {\cal S}} {{x_{ks}}}  + \sum\limits_{i \in {{\cal F}_1}} {{\lambda _i}{N_{ki}}{x_{ki}}},\\
{n_i} &= \sum\limits_{s \in {\cal S}} {({\lambda _i} - 1){x_{is}}}  + {\lambda _i}\sum\limits_{p \in {{\cal F}_2}} {\left(\sum\limits_{s \in {\cal S}} {x_{ip}^s}  - {x_{pi}}{N_{pi}}\right)}.
\end{align*}

Here, we define a path as a sequence of nodes that originates from some BAN and ends at some SBS. Path $P=\{k,i_1,\ldots,i_q\}$ indicates that BAN $k$ (hop 0) serves SBS $i_1$ and SBS $i_n$ (hop $n$) relays data to $i_{n+1}$ with $1\le n \le (q-1)$. For SBS $i$, $P_i$ and $h_i$ are defined to denote its path and the hop number, respceively. The following parameters are defined to facilitate description of proposed algorithms.
\vspace{-5pt}
\begin{align*}
&{ P}_i^{n}=\{i'|P_{i'}=P_i,h_{i'}=n\}, \\
&{\cal P}_{i,b}=\{i'|i' \in {\cal F}_2, P_{i'}=P_i,h_{i'}<h_i\}, \\
&{\cal P}_{i,a}=\{i'|i' \in {\cal F}_2, P_{i'}=P_i,h_{i'}>h_i\},\\
&L_{P_i}=\mathop {\max} \{h_{i'}|i' \in {\cal F}_2, P_{i'}=P_i\}, \\
&r_i=\sum\limits_{s \in {\cal S}}{x_{is}}.
\end{align*}
\vspace{-5pt}
Then $n_i$ can be replaced by
\[
n_i'=\left(\sum\limits_{p \in {\cal P}_{i,b}}{{\lambda}_p}+{\lambda}_i-1\right)r_i-{\lambda}_iN_{P_i^{(h_i-1)}i}.
\]

Since ($\textrm{P}_{\bm{\lambda}}^{\epsilon}$) is a relaxation of ($\textrm{P}_{\epsilon}$), a lower bound for ($\textrm{P}_{\epsilon}$) can be obtained through solving ($\textrm{P}_{\bm{\lambda}}^{\epsilon}$). However, problem ($\textrm{P}_{\bm{\lambda}}^{\epsilon}$) is still no less complex than the NP-Complete knapsack problem since the benefits of BSs, $m_k$ and $n_i$, depend on each other closely. If we relax constraint (\ref{HopConstraint}) further, the obtained optimal value of ($\textrm{P}_{\bm{\lambda}}^{\epsilon}$) would degrade significantly. Here, we propose a local search algorithm to obtain the best connection variables $\bm x$ and optimal objective value $V(\bm{z},\bm{y},\bm{w})$ given deployment variables $[\bm{z},\bm{y},\bm{w}]$ which is described in Algorithm 1. 

\begin{table}[t] 
\begin{tabular}
{p{0.95\columnwidth}}
\\
\toprule
\textbf{Algorithm 1} Local search algorithm for problem ($\textrm{P}_{\bm{\lambda}}^{\epsilon}$) with fixed BSs.\\
\midrule
\textbf{Input:} $\left[{{\bm{z}},\bm{y}},{\bm{w}}\right]$, $\theta$, ${\lambda}_i$, $N_b$, $N$, $N_{ki}$, $N_{ip}$, $d_m$, $D_{kt}$, $D_{it}$, $D_{jt}$ ($k \in {{\cal F}_1}, i,p \in {{\cal F}_2}, j \in {{\cal F}_3}, m \in {\cal M}$).
\begin{enumerate}[  1:]
\item initialize $\bm{x}=\bm{0}$, ${\cal K}=\{k|z_k=1\}$, ${\cal I}=\{i|y_i=1\}$, ${\cal W}=\{j|w_j=1\}$, $P_{i}^n=0$, ${\cal P}_{i,b}=\emptyset$, ${\cal P}_{i,a}=\emptyset$, $L_{P_i}=0$, $r_i=0$, ${\cal I}_o=\emptyset$.
\item for $s$, ${\cal K}_s=\{k|z_k=1,d_{ks} \le D_{kt}\}$, $k=\mathop{\arg\min}\limits_{k' \in {{\cal K}_s}}{d_{k's}}$, $x_{ks}=1$.
\item \textbf{while (${\cal W}\neq\emptyset$)}
\end{enumerate}
\begin{enumerate}[  1:\quad]
\setcounter{enumi}{3}
\item for $j\in {\cal W}$, $k \in {\cal K}$, denote the maximum coverage of $j$ when $x_{kj}=1$ as $\textrm{Cov}_{kj}$.
\item $\{j_0,k_0\}=\mathop{\arg\max}\limits_{j\in {\cal W}, k \in {\cal K}}{\textrm{Cov}_{kj}}$, $x_{k_0j_0}=1$, update $x_{j_0m}$, remove MA $j_0$ from $\cal W$.
\item if $\sum\limits_{j \in {{\cal F}_3}}{x_{k_0j}}=N_b$, remove BAN $k_0$ from $\cal K$.
\end{enumerate}
\begin{enumerate}[  1:]
\setcounter{enumi}{6}
\item \textbf{end while}
\item compute the objective value of ($\textrm{P}_{\bm{\lambda}}^{\epsilon}$) and denote as $V$.
\item \textbf{while (${\cal I}\neq\emptyset$)}
\end{enumerate}
\begin{enumerate}[  1:\quad]
\setcounter{enumi}{9}
\item for $i \in {\cal I}$, $k \in {\cal K}$, denote the maximum coverage of $i$ when $x_{ki}=1$ as $\textrm{Cov}_{ki}$, denote $\Delta{V_{ki}}=-\textrm{Cov}_{ki}$.
\item for $i \in {\cal I}$, $p \in {{\cal I}_o}$, if $L_p<N+1$, denote $\Delta{V_{ip}}$=\{Objective value change when $P_p^{h_{p}-1}=i$\} and $\Delta{V_{pi}}$=\{Objective value change when $P_p^{h_{p}+1}=i$\}.
\item $\{i_0\}=\mathop{\arg\min}\limits_{i \in {\cal I}}\{\Delta{V_{ki}},\Delta{V_{pi}},\Delta{V_{ip}},k \in {\cal K},p \in {{\cal I}_o}\}$.
\item if $\Delta{V_{k_0i_0}}=\Delta V=\mathop{\min}\limits_{i \in {\cal I}}\{\Delta{V_{ki}},\Delta{V_{pi}},\Delta{V_{ip}},k \in {\cal K},p \in {{\cal I}_o}\}$, $x_{k_0i_0}=1$, update path $P_{i_0}$ and $L_{P_{i_0}}$. if $\sum\limits_{i \in {{\cal I}_o}}{x_{k_0i}}+\sum\limits_{j \in {{\cal F}_3}}{x_{k_0j}}=N_b$, remove BAN $k_0$ from $\cal K$.
\item if ${\Delta{V_{p_{0}i_0}}}=\Delta V$, insert $i_0$ between $p_0$ and $P_{p_0}^{h_{p_0}-1}$; if $\Delta{V_{i_0p_{0}}}=\Delta V$, insert $i_0$ between $p_0$ and $P_{p_0}^{h_{p_0}+1}$. $P_{i_0}=P_{p_0}$, update path $P_{i_0}$.
\item remove $i_0$ from $\cal I$ and ${\cal I}_o={{\cal I}_o}\cup\{i_0\}$, $V=V+\Delta V$.
\end{enumerate}
\begin{enumerate}[  1:]
\setcounter{enumi}{15}
\item \textbf{end while}
\end{enumerate}
\textbf{Output:} $\bm{x}$, $V(\bm{z},\bm{y},\bm{w})=V$\\
\bottomrule
\end{tabular}
\vspace{-12pt}
\end{table}

In line 11 in the Algorithm 1, for SBS $i$, we provide the following procedure to compute the $\Delta{V_{ip}}$.
\begin{enumerate}[Step 1:]
\item $\Delta{V_{ip}}=0$. For SBS $p' \in {{\cal P}_{p,a}}$, if $\sum\limits_{i' \in {{\cal P}_{p',b}}}{{\lambda}_{i'}}+{\lambda}_i>1$, $\Delta{V_{ip}}=\Delta{V_{ip}}-(\sum\limits_{i' \in {{\cal P}_{p',b}}}{{\lambda}_{i'}-1)r_{p'}}$ and $r_{p'}=0$.
\item For those $p'$ with $r_{p'}=0$, $x_{p's}=0$ for $s \in {\cal S}$.
\item If $\sum\limits_{i' \in {{\cal P}_{p,b}}}{{\lambda}_{i'}}+{\lambda}_i>1$, $r_i=0$.  If $\sum\limits_{i' \in {{\cal P}_{p,b}}}{{\lambda}_{i'}}+{\lambda}_i\le 1$, based on the updated connection variables in step 2, determine $r_i$ of SBS $i$ when it is connected to $P_{p}^{h_p-1}$. $\Delta{V_{ip}}=\Delta{V_{ip}}+(\sum\limits_{i' \in {{\cal P}_{p,b}}}{{\lambda}_{i'}}+{\lambda}_i-1)r_i$.
\item If $P_{p}^{h_p-1}$ is BAN $k_0$, $\Delta{V_{ip}}=\Delta{V_{ip}}+{\lambda}_p (N_{k_0p}-N_{ip})-{\lambda}_i N_{k_0i}$.
\item If $P_{p}^{h_p-1}$ is SBS $i_0$, $\Delta{V_{ip}}=\Delta{V_{ip}}+{\lambda}_p(N_{i_0p}-N_{ip})-{{\lambda}_i}N_{i_0i}$.
\end{enumerate}
$\Delta{V_{pi}}$ can be calculated in a similar way. Based on Algorithm 1, next, we propose a tabu search based algorithm to solve ($\textrm{P}_{\bm{\lambda}}^{\epsilon}$).
\subsection{Tabu search}
Tabu search is a local search scheme that can avoid being trapped in local optimal solutions through a short term memory called tabu list.
Here, we propose a tabu search algorithm to solve ($\textrm{P}_{\bm{\lambda}}^{\epsilon}$) and obtain Pareto optimal solution set for (P).

Here, we define three local moves to search the neighborhood ${\cal N}(\bm{z},\bm{y},\bm{w})$ of a solution $[\bm{z},\bm{y},\bm{w}]$: 1) \emph{Open move}: deploy a station at an empty site; 2) \emph{Close move}: remove a deployed station; 3) \emph{Swap move}: remove a deployed station and deploy a new station. It needs to be noted that the objective values of solutions in ${\cal N}(\bm{z},\bm{y},\bm{w})$ can not be obtained from the objective value of $[\bm{z},\bm{y},\bm{w}]$. Optimal variables $\bm{x}$ given deployment variables can only be got through solving the problem.

To solve ($\textrm{P}_{\bm{\lambda}}^{\epsilon}$) first, we propose a two-level tabu search algorithm described in Algorithm 2 to get the optimal solution $[\bm{z},\bm{y},\bm{w},\bm{x}]$ and optimal value $V_B=V(\bm{z},\bm{y},\bm{w})$. In Algorithm 2, restart diversification scheme is used: re-deploy $N_{div}$ rarely deployed stations when there is no non-tabu solutions in the neighborhood. Meanwhile, the two-level structure acts as an intensification scheme to search the solution space of SBSs and MAs thoroughly. In Algorithm 2, the Algorithm 1 is used in each iteration, line 9 and 14, to calculate the objective value $V(\bm{z}',\bm{y}',\bm{w}')$ for each solution $(\bm{z}',\bm{y}',\bm{w}') \in {\cal N}(\bm{z},\bm{y},\bm{w})$.

Given the output of the Algorithm 2, $[\bm{z},\bm{y},\bm{w}]$, a two-level search algorithm, as included in Algorithm 3 from line 9 to line 19, is executed to obtain Pareto optimal solution set for problem (P). Unlike Algorithm 2, multiple Pareto optimal solutions rather than one optimal solution can be got. In the algorithm, a new parameter $\Delta{\epsilon}$ is introduced to control the intensification scheme. In each iteration, only solution space with cost in $[\epsilon-\Delta\epsilon,\epsilon]$ are searched. It needs to be noted that in the search of solutions for problem (P), objective functions ${f_1}\left(\bm{z},\bm{y},\bm{w}\right)$ and $f_c({\bm x})$ are used to determine whether a solution is efficient or not.
\begin{table}[t] 
\begin{tabular}
{p{0.95\columnwidth}}
\\
\toprule
\textbf{Algorithm 2} Algorithm for problem (${\textrm{P}}_{\bm{\lambda}}^{\epsilon}$).\\
\midrule
\textbf{Input:}  $N_{t_1},N_{t_2}$, $N_{swap}$, $N_{div}$, tabu list ${\cal T}_1$ and ${\cal T}_2$, $\epsilon$, $\theta$, ${\lambda}_i$, $N_b$, $N$, $N_{ki}$, $N_{ip}$, $d_m$, $D_{kt}$, $D_{it}$, $D_{jt}$ ($k \in {{\cal F}_1}, i,p \in {{\cal F}_2}, j \in {{\cal F}_3}, m \in {\cal M}$).
\begin{enumerate}[  1:]
\item initialize $t_1=0,t_2=0$, empty tabu list, $[\bm{y},\bm{z},\bm{w}]=\bm{0}$.
\item \textbf{while (${f_1}\left(\bm{z},\bm{y},\bm{w}\right) < {\epsilon}$)}
\end{enumerate}
\begin{enumerate}[  1:\quad]
\setcounter{enumi}{2}
\item $k=\mathop{\arg\min}\limits_{k' \in {\cal F}_1, z_{k'}=0 }{c_k}$. if ${f_1}\left(\bm{z},\bm{y},\bm{w}\right)+c_{k} \le \epsilon$, $z_k=1$.
\item if ${f_1}\left(\bm{z},\bm{y},\bm{w}\right)+c_{k} > \epsilon$ or $\bm{z}=\bm{1}$, $p=\mathop{\arg\min}\limits_{p' \in {{\cal F}_2}\cup{{\cal F}_3}, y_{p'}=0, w_{p'}=0  }{c_k}$.
\item if ${f_1}\left(\bm{z},\bm{y},\bm{w}\right)+c_{p} \le \epsilon$, open facility $p$; else, break.
\end{enumerate}
\begin{enumerate}[  1:]
\setcounter{enumi}{5}
\item \textbf{end while}
\item Given $[\bm{z},\bm{y},\bm{w}]$, get objective value $V_B=V(\bm{z},\bm{y},\bm{w})$ through Algorithm 1. $\bm{v}_b=[\bm{z},\bm{y},\bm{w}]$.
\item \textbf{while ($t_1<N_{t_1}$)}
\end{enumerate}
\begin{enumerate}[  1:\quad]
\setcounter{enumi}{8}
\item compute ${\cal N}(\bm{z})$ with fixed $[\bm{y},\bm{w}]$ and objective value through Algorithm 1, $\bm{p}=\mathop{\arg\min}\limits_{\bm{z}' \in {\cal N}(\bm{z})}{V(\bm{z}',\bm{y},\bm{w})}$
\item if ${V(\bm{p},\bm{y},\bm{w})}<V_B$; $V_B=V(\bm{p},\bm{y},\bm{w})$, $\bm{v}_b=[\bm{p},\bm{y},\bm{w}]$.
\item if ${V(\bm{p},\bm{y},\bm{w})}\ge V_B$, $\bm{p}=\mathop{\arg\min}\limits_{\bm{z}' \in {\cal N}(\bm{z}), \bm{z}' \textrm{ is non-tabu}}{V(\bm{z}',\bm{y},\bm{w})}$.
\item $\bm{z}=\bm{p}$, $t_1=t_1+1$, update ${\cal T}_1$.
\item \textbf{while ($t_2<N_{t_2}$)}
\end{enumerate}
\begin{enumerate}[  1:\qquad]
\setcounter{enumi}{13}
\item compute ${\cal N}(\bm{y},\bm{w})$ with fixed $\bm{z}$ and objective value through Algorithm 1, $\bm{q}=\mathop{\arg\min}\limits_{\bm{q}' \in {\cal N}(\bm{y},\bm{w})}{V(\bm{z},\bm{q})}$
\item if ${V(\bm{z},\bm{q})}<V_B$; $V_B=V(\bm{z},\bm{q})$, $\bm{v}_b=[\bm{z},\bm{q}]$.
\item if ${V(\bm{z},\bm{q})}\ge V_B$, $\bm{q}=\mathop{\arg\min}\limits_{\bm{q}' \in {\cal N}(\bm{y},\bm{w})), \bm{q}' \textrm{ is non-tabu}}{V(\bm{z},\bm{q})}$.
\item if any $\bm{q}' \in {\cal N}(\bm{y},\bm{w})$ is tabu, deploy $N_{div}$ rarely deployed facilities and clear ${\cal T}_2$; else,     $[\bm{y},\bm{w}]=\bm{q}$, update ${\cal T}_2$. $t_2=t_2+1$.
\end{enumerate}
\begin{enumerate}[  1:\quad]
\setcounter{enumi}{17}
\item \textbf{end while}
\item $t_2=0$
\end{enumerate}
\begin{enumerate}[  1:]
\setcounter{enumi}{19}
\item \textbf{end while}
\item get $\bm x$ and $V(\bm{z},\bm{y},\bm{w}$)  with $\bm{v}_b=[\bm{z},\bm{y},\bm{w}]$ through Algorithm 1.
\end{enumerate}
\textbf{Output:} $[\bm{z},\bm{y},\bm{w},\bm{x}]$ and $V_B=V(\bm{z},\bm{y},\bm{w})$, $\bm{v}_b=[\bm{z},\bm{y},\bm{w}]$. \\
\bottomrule
\end{tabular}
\vspace{-14pt}
\end{table}
\subsection{Update of $\epsilon$}
In order to get the entire optimal solution set, $\epsilon$ needs to be decreased gradually. All nondominated solutions can be obtained through the update of $\epsilon$ \cite{Kirlik14}. Given ${\epsilon}_t$ in iteration $t$, optimal solution set ${\cal P}_{{\epsilon}_t}$ can be obtained through the proposed algorithm. Then, ${\epsilon}_t$ can be updated adaptively as follows:
\[
{\epsilon}_{t+1}={\min}\left(\mathop {\min }\limits_{\left[{\bm{z}'},{\bm{y}'},{\bm{w}'}\right] \in {{\cal{P}}_{{\epsilon}_t}}} {f_1}({\bm{z}'},{\bm{y}'},{\bm{w}'}), {\epsilon}_t\right) -\Delta c.
\vspace{-3pt}
\]
When ${\cal P}_{{\epsilon}_t}=\emptyset$, we update the cost constraint based on ${\epsilon}_t$. $\Delta c$ is a small positive number.

The algorithm to solve (P) is summarized as Algorithm 3.

\begin{table}[t]
\begin{tabular}
{p{0.95\columnwidth}}
\\
\toprule
\textbf{Algorithm 3} Algorithm to solve (P).\\
\midrule
\textbf{Input:} $N_b$, $N_{max,L}$, $\Delta\epsilon$, $N_{t_1,max}$, $N_{t_2,max}$.
\begin{enumerate}[  1:]
\item initialize ${\epsilon}_0=\sum\limits_{k \in {\cal{F}}_1}{c_k}+\sum\limits_{i \in {\cal{F}}_2}{c_i}+\sum\limits_{j \in {\cal{F}}_3}{c_j}$, $\cal{P}=\emptyset$, $t=0$, $t_L=0$, ${{\cal B}_l}=\emptyset$.
\item \textbf{while (${\epsilon}_t > {\mathop{\min}\limits_{k \in {{\cal F}_1}}}{c_k}$)}
\end{enumerate}
\begin{enumerate}[  1:\quad]
\setcounter{enumi}{2}
\item initialize $\bm{\lambda}$, ${\cal P}_r=\emptyset$, ${{\cal{V}}}_r=\emptyset$,$t_L=0$, $t_1=0$, $t_2=0$.
\item \textbf{while ($t_L < N_{max,L}$)}
\end{enumerate}
\begin{enumerate}[  1:\qquad]
\setcounter{enumi}{4}
\item solve (${\textrm{P}}_{\bm{\lambda}}^{{\epsilon}_t}$) through Algorithm 2, ${\cal P}_r={\cal P}_r \cup \{V_B\}$, ${{\cal{V}}}_r={{\cal{V}}}_r \cup\{{\bm{v}}_b\}$.
\item $t_L=t_L+1$, update $\bm{\lambda}$.
\end{enumerate}
\begin{enumerate}[  1:\quad]
\setcounter{enumi}{6}
\item \textbf{end while}
\item $[\bm{z},\bm{y},\bm{w}]=\mathop{\arg\max}\limits_{[\bm{z}',\bm{y}',\bm{w}'] \in {{\cal V}_r}}{V(\bm{z}',\bm{y}',\bm{w}')}$, ${\cal B}_l={\cal B}_l \cup \{V({\bm{z},\bm{y},\bm{w}})\}$
\item \textbf{while ($t_1<N_{t_1,max}$)}
\end{enumerate}
\begin{enumerate}[  1:\qquad]
\setcounter{enumi}{9}
\item compute ${\cal N}(\bm{z})$ with fixed $[\bm{y},\bm{w}]$, ${\cal P}_e=\{[\bm{z}',\bm{y},\bm{w}]|\bm{z}' \in {{\cal N}({\bm z}}),{f_1}\left(\bm{z}',\bm{y},\bm{w}\right) \in [\epsilon-\Delta\epsilon,\epsilon],[\bm{z}',\bm{y},\bm{w}]$ is nondominated\}.
\item ${\cal P}={\cal P} \cup {{\cal P}_e}$, delete dominated solutions in $\cal P$.
\item $\bm{z}$=\{best non-tabu solution in ${\cal P}_e$\}, $t_1=t_1+1$, update ${\cal T}_1$.
\item \textbf{while ($t_2<N_{t_2,max}$)}
\end{enumerate}
\begin{enumerate}[  1:\qquad\quad]
\setcounter{enumi}{13}
\item compute ${\cal N}(\bm{y},\bm{w})$ with fixed $\bm{z}$, ${\cal P}_e=\{[\bm{z},\bm{y}',\bm{w}']|[\bm{y}',\bm{w}'] \in {{\cal N}({\bm y}},\bm{w}),{f_1}\left(\bm{z},\bm{y}',\bm{w}'\right) \in [\epsilon-\Delta\epsilon,\epsilon],[\bm{z},\bm{y}',\bm{w}']$ is nondominated\}.
\item ${\cal P}={\cal P} \cup {{\cal P}_e}$, delete dominated solutions in $\cal P$.
\item $[\bm{y},\bm{w}]$=\{best non-tabu solution in ${\cal P}_e$\}, update ${\cal T}_2$. $t_2=t_2+1$.
\end{enumerate}
\begin{enumerate}[  1:\qquad]
\setcounter{enumi}{16}
\item \textbf{end while}
\item $t_2=0$
\end{enumerate}
\begin{enumerate}[  1:\quad]
\setcounter{enumi}{18}
\item \textbf{end while}
\item $t_1=0$, $t=t+1$, update ${\epsilon}_t$.
\end{enumerate}
\begin{enumerate}[  1:]
\setcounter{enumi}{20}
\item \textbf{end while}
\end{enumerate}
\textbf{Output:} Nondominated solution set $\cal{P}$,  a lower bound set ${\cal{B}}_{l}$. \\
\bottomrule
\end{tabular}
\vspace{-5pt}
\end{table}
\section{Simulation results and discussions}
In our simulation, a 73GHz band is considered by exploiting the measurements in \cite{Ghosh14}. $\beta$ is set as 0.046 \cite{Kulkarni14}.
For users, the SNR threshold, ${\gamma}_t$, is -10dB calculated from a 100Mbps outage rate with a 1GHz bandwidth \cite{Ghosh14}. The bandwidth of the backhaul link is also 1GHz and the noise power is ${\sigma}^2=$-74dBm \cite{Kulkarni14} for mmW band. The access and backhaul outage probability are set 0.1. The subarea is  set as $10\textrm{m}\times10\textrm{m}$ and the density of users is 200/${\textrm{km}}^2$. The maximum number of relays for an SBS, $N$, is set as 2 and the backhaul capability of a BAN, $N_b$, is set as 5. $\theta$ is set to 0.5. For MTC, the results in \cite{Zhou13} is used where an MA can support 600 machines with a latency requirement of 15 ms.

\begin{figure}[!t]
\setlength{\abovecaptionskip}{3pt}
\setlength{\belowcaptionskip}{0pt}
\centering
\scalebox{0.39}{\includegraphics*[11,5][493,375]{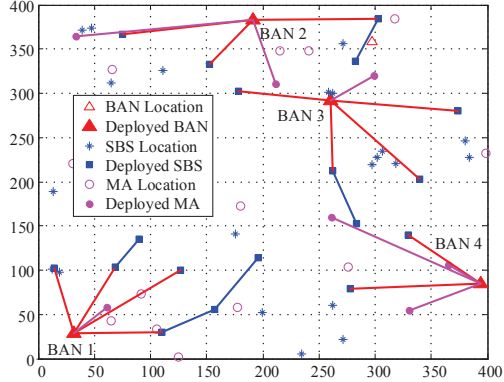}}
\caption{Deployment instance when the number of potential locations for BANs, SBSs and MAs is 5, 40 and 20, respectively.}
\vspace{-12pt}
\label{DeploymentInstance}
\end{figure}

Fig. \ref{DeploymentInstance} shows a deployment instance which achieves the minimum objective $f_c(\bm x)$, \emph{best} coverage, in a 400m $\times$ 400m area. In this area, 2000 machines are distributed uniformly. Here, the deployed BANs, SBSs, MAs and connection relationships are shown in the figure. Better coverage has been obtained through the proposed framework compared with conventional fiber-based and single-hop wireless backhaul models. In the result, 4 BANs have been deployed which provide backhaul links to 18 SBSs and 7 MAs. One potential site for BANs has not been used since the SBSs in the 2-hop wireless backhaul path of BAN 2 have covered the subareas around this potential site. Thus, through the utilization of sites without fiber connection, less BANs are needed compared with conventional models.


%

\begin{figure}[!t]
\setlength{\abovecaptionskip}{3pt}
\setlength{\belowcaptionskip}{0pt}
\centering
\scalebox{0.39}{\includegraphics*[5,2][512,400]{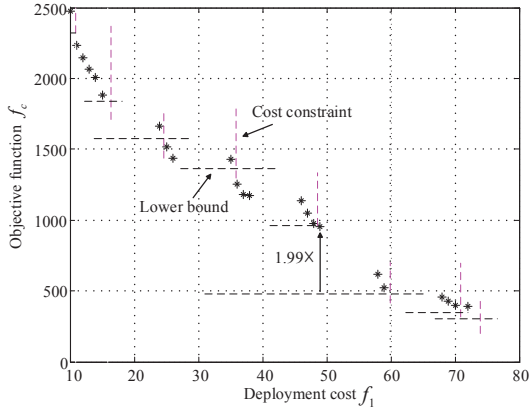}}
\caption{Obtained Pareto optimal solutions and lower bounds.}
\vspace{-14pt}
\label{OptimalSolutions}
\end{figure}

In Fig. \ref{OptimalSolutions}, we show the Pareto optimal solutions and lower bounds obtained using the proposed algorithm for an 400m $\times$ 400m area with 5 locations, 40 locations and 20 locations for BANs, SBSs and MAs, respectively. Here, the deployment costs of an SBS and an MA is normalized to 1 and the deployment cost of a BAN is 10. Fig. \ref{OptimalSolutions} shows that the gap between the obtained solutions and the lower bound is quite tight. The maximum ratio of the objective value of the optimal solutions to the lower bound is 1.99. Clearly, the proposed algorithm is quite effective through the utilization of the structure of the problem.

\section{Conclusions}
In this paper, we have proposed a general deployment framework for next-generation cellular networks which incorporates HTC, MTC, wired backhaul links and multi-hop wireless backhaul links. A general multi-objective integer programming model has been proposed to determine the location of BANs, SBSs, MAs and their connections. To obtain the Pareto optimal solution set, we have proposed an adaptive heuristic algorithm by combining the weighted sum method, the  $\epsilon$-method, Lagrangian relaxation and tabu search jointly. The proposed algorithm can obtain both solutions and lower bounds which can be used to evaluate the quality of the obtained solutions. Simulation results have shown that the proposed framework can provide better deployment solutions  than conventional fiber-based and single-hop wireless backhaul model. Furthermore, the objective values of obtained solutions are quite close to the lower bounds.

\small
\setlength{\baselineskip}{0.95\baselineskip}
\def\baselinestretch{0.78}
\footnotesize
\def\baselinestretch{0.78}
\bibliographystyle{IEEEtran}
\bibliography{refs_paper}

\end{document}